\documentstyle[11pt,newpasp,twoside,epsf]{article}
\def\edcomment#1{\iffalse\marginpar{\raggedright\sl#1\/}\else\relax\fi}
\reversemarginpar

\begin{document}

\title{ The impact of dust and ionization effects on abundance measurements
of Damped Ly$\alpha$ systems }
\author{Giovanni Vladilo}
\affil{Osservatorio Astronomico di Trieste, Via G.B. Tiepolo 11, \\ 
34131 Trieste, Italy}
 
\begin{abstract}
Studies of elemental abundances  are a fundamental tool
for unveiling the nature of the  high-redshift (proto-)galaxies
associated to  Damped Ly $\alpha$ systems (DLAs).  
The present contribution 
analyses the impact of 
dust and ionization effects on abundance measurements in DLAs.
The behaviour of the $\alpha$/Fe abundance ratio corrected for such effects 
is  used to derive information on the chemical history 
and nature of DLA galaxies. 
The $\alpha$/Fe data indicate that DLAs at $z \approx 2.5$ do not represent
a homogeneous class of objects. On average, DLAs show
non-enhanced $\alpha$/Fe ratios at  low metallicity, suggesting an  
 origin  in galaxies with low or intermittent star formation rates.  
\end{abstract}

\section{Introduction}

Damped Ly\,$\alpha$ systems (DLAs) have the highest HI column densities
among QSO absorbers ($N$(HI) $\geq 2 \times 10^{20}$ atoms cm$^{-2}$) and 
contain the bulk of neutral gas observed at $z \approx 2.5$. 
The first studies of the baryonic content   in DLAs  suggested
that their co-moving mass density,  $\Omega_{\rm DLAs}(z)$, increases with $z$ and that
$\Omega_{\rm DLAs}(z \approx 2.5)$  is  comparable to 
the baryonic density of visible matter in local galaxies, $\Omega_{\rm stars}(z=0)$
(Wolfe et al. 1995).   
These results lead to the proposal that DLAs represent the  gas of
high redshift (proto-)galaxies which
is consumed  in the
course of the cosmic evolution because it is transformed into stars. 
Recent results do not confirm the evolution of  $\Omega_{\rm DLAs}(z)$ 
(Rao \& Turnshek 2000)
 and indicate that $\Omega_{\rm DLAs} (z \simeq 2.5) < \Omega_{\rm stars}(z=0)$ 
(Storrie-Lombardi \& Wolfe 2000, P\'eroux et al. 2001). 
However, there is still a general consensus,
based on photometric and spectroscopic observations,
that DLAs originate in the diffuse gas of (proto-)galaxies. 
One of the main goals of DLA research is  to understand the role of the
associated galaxies
 in  the   general context of the high redshift universe.  
At $z \la 1$ photometric studies 
indicate that  the candidate DLA galaxies  have 
different morphologies and luminosities, with spirals being a small fraction of
the sample (Le Brun et al. 1997, Turnshek et al. 2001 and refs. therein).
At higher redshift the photometric identification of the intervening galaxies 
 is more difficult 
and only preliminary results are currently available
(Warren et al. 2001).
Spectroscopy is the most powerful tool for understanding the nature of DLAs at
$z \geq 2$,   the range where   DLAs can be  identified and investigated
with ground based facilities.
In particular, spectroscopy   has been used to 
study  the kinematics and the abundances.
The kinematical data are useful for probing semi-analytical 
models of galaxy formation in the context of different
cosmogonies, but are not able to distinguish between  
an origin of DLAs in massive disks or in low-mass proto-galactic clumps
(Prochaska \& Wolfe 1997;
Haehnelt, Steinmetz \& Rauch 1998;
Wolfe \& Prochaska 2000b).
The elemental abundances can be used to probe the chemical evolution
state of DLAs   
from the comparison with the abundances observed in   
metal-poor astrophysical sites  and with
the   predictions of galactic evolution models. 
However, 
the interpretation of DLAs abundances can lead to rather different conclusions
depending on whether dust and ionization corrections 
are considered to be important or not. For instance,  
 the overabundances of the Si/Fe ratio measured in DLAs
have been ascribed 
to nucleosynthetic processes (Lu et al. 1996), to differential dust depletion
(Vladilo 1998; hereafter V98), and to ionization effects
(Izotov, Schaerer \& Charbonnel 2001).   
The present contribution summarizes the studies 
  of ionization and dust effects in DLAs
(Sections 2 and 3, respectively) and 
the behaviour of the $\alpha$/Fe
 ratio  corrected for such effects (Section 4).

\section{Ionization effects}

The majority of the metal species observed in DLAs is in the ionization
state typical of interstellar HI regions.
In such regions most of the elements are singly ionized  
(e.g. C\,II, Mg\,II, Al\,II, Si\,II, S\,II, Cr\,II, Mn\,II, Fe\,II, Ni\,II and Zn\,II)  
and only those   with   ionization potential
 IP$_{\rm I}$  $>$ 13.6 eV are in the neutral state
(e.g.  N\,I, O\,I, and Ar\,I). 
This can be understood assuming that HI regions are   
exposed to a radiation field, the photoionization   
of species with IP $< 13.6$ eV 
being more efficient than competing processes. 
%
%
%
Column density studies suggest that these conditions hold in DLAs where,
for instance,  
$N$(Mg\,II) $\gg$  $N$(Mg\,I), $N$(C\,II) $\gg$ $N$(C\,I), and $N$(Al\,II) $>$ $N$(Al\,III)
(Lu et al. 1996, Prochaska \& Wolfe 1999). 
Studies of 
the absorption profiles indicate that the typical HI species
listed above have very similar 
radial velocity distributions, consistent with a common
origin in HI regions  (Lu et al. 1996, Wolfe \& Prochaska 2000a). 
Taken together, these results justify the habit of 
deriving abundances of DLAs from the species of low ionization without
applying ionization corrections. 
In fact,  early computations of the ionization balance in DLAs confirmed that 
ionization effects are, in general, small or negligible  
(Viegas 1995; Lu et al. 1995; Prochaska \& Wolfe 1996).
However, the observations also  reveal species in higher states
of ionization at the same redshift of the DLAs,
such as Al~III, C\,IV and Si\,IV. 
The  radial velocity profiles 
suggest  that C\,IV and Si\,IV originate in  
volumes of space physically distinct from those occupied by HI regions
(Wolfe \& Prochaska 2000a). 
The C\,IV/Si\,IV regions   are not expected to affect
 abundance determinations since they do  not contribute 
to the column densities  of the HI   species. 
The radial velocity profiles of
Al~III  suggest instead a common origin with the species of lower ionization
(Wolfe \& Prochaska 2000a). 
This fact is not easy to understand in the framework of the simple 
model of a   HI region exposed to  ionizing radiation  
and has stimulated  
new investigations on the ionization balance of DLAs
(Howk \& Sembach 1999; Izotov et al. 2001;
Vladilo et al. 2001; hereafter VCBH01). 
%
%
Considering that  Al~III requires photons with $h\nu > 18.8$ eV
to be produced,
two  hypothesis for its origin can be envisaged:

\begin{description}

\item
(1)
The Al~III arises in the same region where the typical HI species 
are located. This can be the case if the ionizing continuum 
is hard and there are enough
high energy photons that can leak into the neutral region 
(the HI photoionization cross-section  declines as $\nu^{-3}$).
Ionization models built in the framework of this hypothesis
are  characterized by a single region embedded in
a hard continuum (called hereafter "1H models").

\item
(2)
The Al~III arises in a partially ionized interface at the border of
the HI region. This can happen
if the ionizing continuum is soft, in which case there are not enough
energetic photons that can penetrate the neutral region. 
The neutral region  contains the typical HI species, but not Al~III. 
Ionization models built in the framework of this  hypothesis
are characterized by two regions and a soft continuum ("2S models").  

\end{description}

Whatever the origin of Al~III is, we expect that other species
of moderate ionization, such as Si\,III or Fe\,III, can be similarly produced.
Unfortunately these species are difficult to
detect in DLAs. As a consequence, the ratio $R$(Al~III/Al~II) $\equiv$ 
$N$(Al~III)/$N$(Al~II) is a unique diagnostic tool to  probe models of DLA photoionization. 
Computations  
constrained by the  $R$(Al~III/Al~II) ratio were first performed for two
individual DLAs (Lu et al. 1995; Prochaska \& Wolfe 1996).
In both cases "1H models" were adopted and the ionization effects
were found to be small or negligible. 
In order to assess the general importance of ionization effects 
 it is therefore necessary
to determine $R$(Al~III/Al~II) in  a large number of DLAs and to 
consider "2S models" as well as "1H models".  
Unfortunately, the Al~II lines are often saturated and only a few
accurate determinations of $R$(Al~III/Al~II) exist. 
To bypass this problem VCBH01 estimated $N$(Al\,II) indirectly
from an empirical correlation found to exist between Si\,II, Al\,II, and Fe\,II
column densities. As a result,  
a sample of 20  $R$(Al~III/Al~II) determinations was obtained. 
As shown in Fig. 1 (left panel), 
the ratio is anti-correlated with the HI column density. 
This trend can be easily reproduced by means of "2S models"
computed at constant values of the photoionization parameter $U$
(i.e. the number of ionizing photons per hydrogen atom). 
The trend can also be reproduced by means of "1H models", assuming
that $U$ scales with a law of the type $U \propto N({\rm HI})^{-1.5}$. 
In any case the anti-correlation between 
log  $R$(Al~III/Al~II) and log $N$(HI)  is  a powerful constraint for estimating 
ionization corrections\footnote{
The correction term $C(\mathrm{X/Y})$ 
is defined here in such a way that 
$ ({\mathrm{X/Y}})_c  =  C({\mathrm{X/Y}}) ~
[ N({\mathrm{X}}^{i_d}) / N({\mathrm{Y}}^{i_d}) ] $,
where (X/Y)$_c$ is the abundance corrected for ionization effects and  
$ N({\mathrm{X}}^{i_d}) / N({\mathrm{Y}}^{i_d})$
is the measured column density ratio of the dominant states of ionization. 
}. 
Examples of correction terms estimated in this way are shown in Fig. 1 (right panel). 
The ionization corrections are generally small and tend to become
less and less important with increasing $N$(HI)
no matter if one uses 1H or 2S models. 
It is important to note that 
the ionized interface and the neutral region are taken to
have similar metallicities
in the VCBH01 models. 
Assuming that
the ionized region has been enriched by metals while the neutral region
is essentially free of metals  Izotov et a. (2001) found 
that ionization corrections can be quite large.  
With the possible exceptions of specific DLAs where these particular
conditions might hold, 
 the ionization corrections are  small or
negligible for the elements most commonly measured in DLAs.

\begin{figure} 
\plottwo{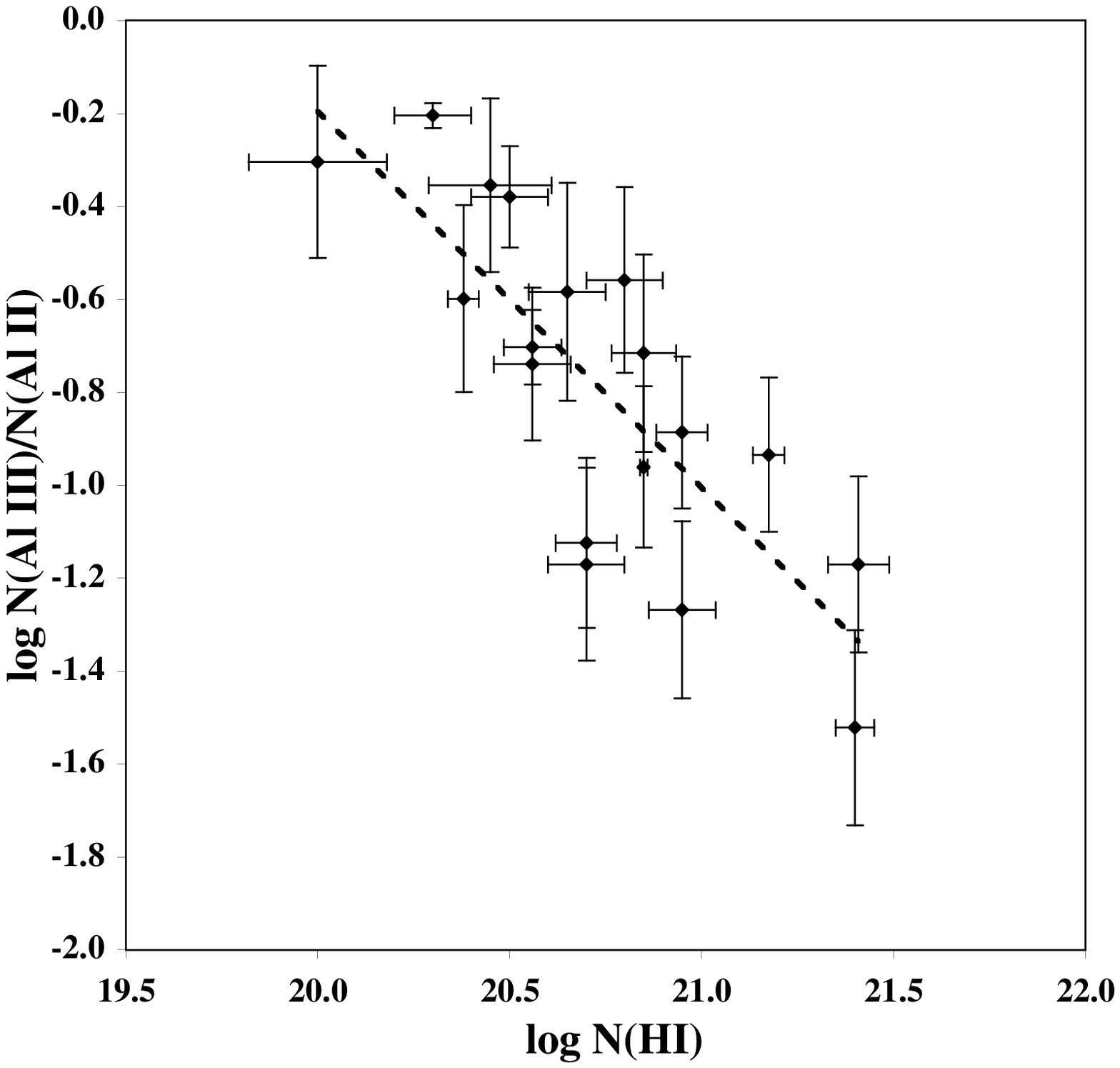}{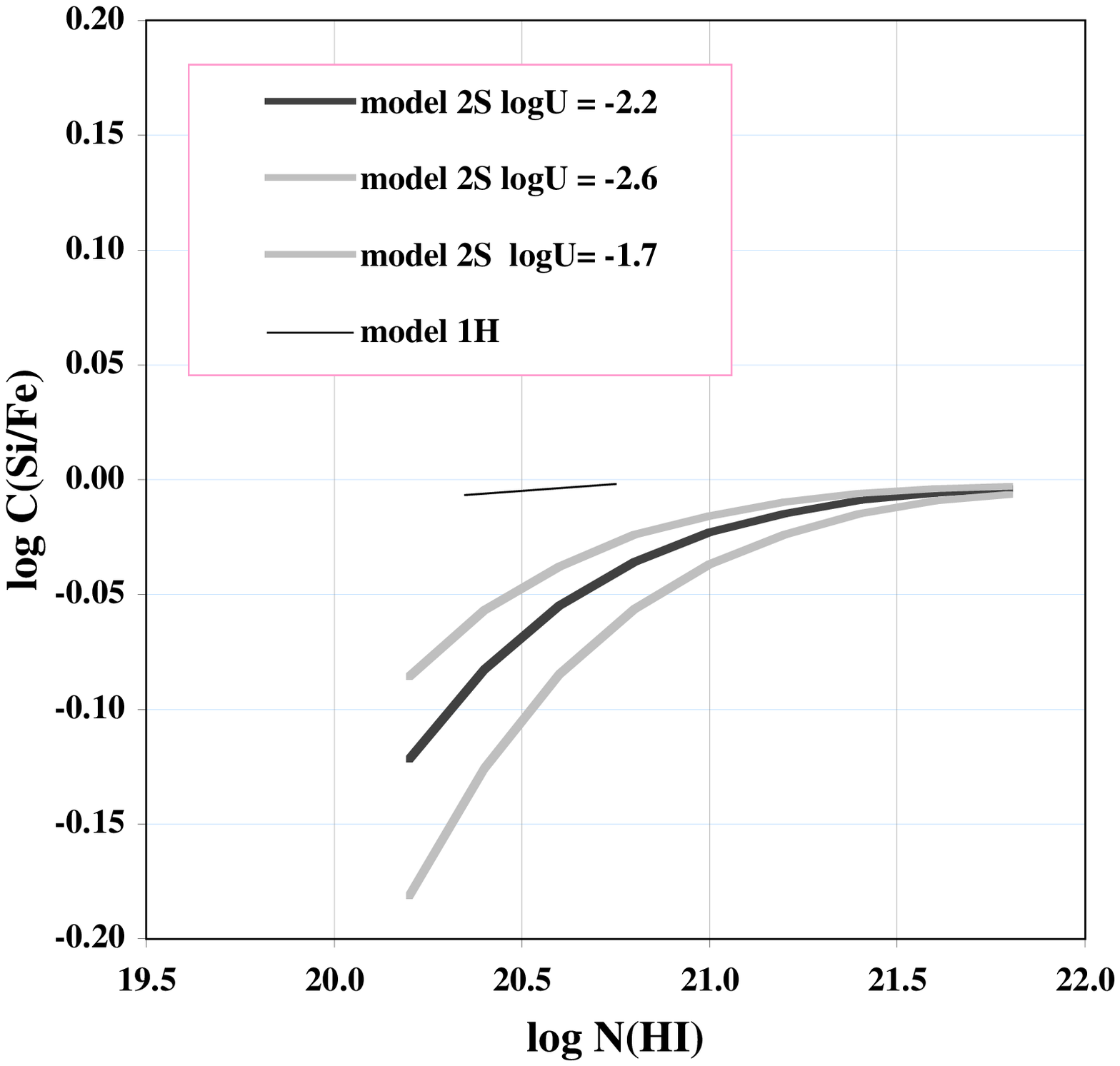}
\caption{
{\bf Left panel}: column density ratio $N$(Al~III)/$N$(Al~II) in DLAs
plotted versus HI column density taken from the compilation by VCBH01. 
The dashed line represents the linear regression through the observed data points. 
{\bf Right panel}:
Ionization correction term for the Si/Fe abundance ratio in DLAs 
predicted by 2S models (thick curves) and  1H models (thin curve) 
computed by VCBH01.
These models have been constrained by the requirement to match the
trend between $N$(Al~III)/$N$(Al~II) and $N$(H\,I)
shown in the left panel.
 }
\end{figure}

\section{Dust effects}

Local interstellar studies indicate that a fraction of the elements
is   not detected in the gas phase because it
is locked into dust grains, an effect referred to as "elemental depletion"
(Savage \& Sembach 1996 and refs. therein). The fraction in dust  
varies for different elements, being close to unity for
refractory elements, and changes in different interstellar environments,
being the highest in cold, dense clouds. 
If dust is present in DLAs we expect elemental depletion to affect the  
abundance determinations, in particular those of refractory elements. 
Two independent types of observations  
suggest that dust is present in DLAs. 
On the one hand,  QSOs with foreground DLAs   
have optical spectral indices different from QSOs without foreground DLAs,
suggestive of reddening due to the intervening absorbers
(Pei, Fall \& Bechtold 1991).
On the other hand, the relative abundances  
Zn/Fe and Zn/Cr in DLAs show significant deviations from 
the corresponding solar values. Such deviations are hard
to   explain  as  intrinsic nucleosynthetic effects and can instead be 
interpreted in terms of differential dust depletion 
(Pettini et al. 1997). 
In addition, Hou et al. (2001) claim that the ratios of refractory elements over Zn
(e.g. Fe/Zn, Cr/Zn, Si/Zn, etc.) 
are anti-correlated with the metal column densities and consider  this result
as an unambiguous sign of dust depletion. 

Different approaches are used to bypass  the problem of dust depletion.
  One is to avoid the use of refractory elements 
(e.g. Fe, Cr, Ni) and to 
use instead elements essentially undepleted,
such as N (Lu et al. 1998, Centuri\'on et al. 1998),
O (Molaro et al. 2000),
S (Centuri\'on et al. 2000), 
and Zn (Pettini et al. 1997, 1999; Vladilo 2000). 
Another approach is to study DLAs apparently free of dust.  
For instance, the $z=3.390$ system toward Q0000-26 does not show any 
evidence of differential Zn/Fe depletion and its  detailed analysis  has allowed 
 the chemical properties of the associated  galaxy to be determined
(Molaro et al. 2000). 
These studies which bypass the depletion problems  do not require
any assumption on the dust properties, but their application 
is limited to some elements and
to particular  DLAs.  
The only way to perform a general study
of DLA abundances is to quantify the  depletion effects. 

A basic approach to estimate the   depletions  is
to compare the abundance patterns observed in DLAs  
with the interstellar depletion patterns typical of our Galaxy.
Studies of this type 
indicate that depletions in DLAs resemble those measured
in the warm gas of the halo or disk  of the Milky Way
(Lauroesch et al. 1996; Kulkarni, Fall \& Truran 1997).  
A similar conclusion was obtained by
Savaglio, Panagia \& Stiavelli (2000), who  derived
the dust-corrected metallicity of individual systems  
assuming that the intrinsic abundance ratios of DLAs are solar. 

A method for
deriving dust-corrected  abundance ratios
of individual systems without making {\em a priori} assumptions on the 
abundance pattern of DLAs  
 was presented by V98.  
In that method the dust was assumed to have 
the same composition as the dust in Galactic warm gas;
the dust-to-gas ratio  
of individual DLAs was then estimated
assuming that the observed overabundances of Zn/Fe 
are entirely due to differential depletion.  
A limitation of that method is that the dust composition is taken to
be constant. In addition, some recent investigations 
indicate that it may be risky to assume [Zn/Fe]=0
(Umeda \& Nomoto 2001 and refs. therein)\footnote{
The usual definition
[X/Y] $\equiv$ log [$N$(X)/$N$(Y)] $-$ log (X/Y)$_{\sun}$
is adopted. 
}.
For these  reasons a refined procedure for dust correction has been 
recently developed  (Vladilo 2001, in preparation). 
A preliminary presentation of such procedure is given in the
rest of this section.    
The fraction in dust of an element X, $f_{\mathrm{X}}$, is allowed to vary 
as a function of the dust-to-metals ratio, $\rho$, and of the intrinsic abundances
of the medium, (X/Y)$_{\mathrm{int}}$,  according to an analytical expression
of the type
\begin{equation}
f_{\mathrm{X}} \propto \rho^{(1+\eta_{\mathrm{x}})} ~ 
10^{(\varepsilon_{\mathrm{x}}-1) \left[ \mathrm{X \over Y} \right]_{\mathrm{int}}} ~,
\end{equation}
where Y is an element used as a reference for   abundance
measurements. 
This expression is a generalization of Eq. (11) given in V98, which represents the 
case $\eta_{\mathrm{x}}=0$ and $\varepsilon_{\mathrm{x}}=0$. 
The parameters $\eta_{\mathrm{x}}$  can be calibrated using 
the depletion patterns observed in the Galaxy, where 
$\left[ \mathrm{X / Y} \right]_{\mathrm{int}}=0$. 
With such a calibration all the typical depletion
patterns of the Milky Way can be successfully
reproduced by only varying $\rho$ (Fig. 2, left panel).  
The parameters $\varepsilon_{\mathrm{x}}$ describe the dependence of the dust composition
on the   composition of the medium (gas plus dust). 
These parameters are expected to be in the range $0 \leq \varepsilon_{\mathrm{x}} \leq 1$
and can potentially be calibrated by studying
depletion patterns in galaxies with non-solar abundances. 
The possibility that the intrinsic Zn/Fe ratio in DLAs differs from the solar
value is considered in a self-consistent way in Eq. (1). 
By following the same logical steps described in V98 it is possible to derive
(i) an expression for estimating $\rho$ given the observed [Zn/Fe] and a guess
of  [Zn/Fe]$_{\rm int}$; (ii) an expression for correcting  
abundance ratios given $\rho$. 
This two-step procedure can be repeated for different values
of $\varepsilon_{\mathrm{x}}$ and [Zn/Fe]$_{\rm int}$ and the corrected
abundances estimated in each case. 
An example of application of this procedure to the [Si/Fe] ratio
in the SMC interstellar gas towards  Sk 108 is shown in Fig.2 (right panel).  
The SMC interstellar ratio is clearly affected by  depletion
since it appears to be much more enhanced than in
 SMC stars. The [Si/Fe] corrected for dust  
closely matches the expected stellar values. 
The [Zn/H] ratio is in good agreement with the SMC stellar metallicities,
while the [Fe/H] ratio badly underestimates the metallicity.   
This example illustrates 
the capability of the dust correction procedure to recover 
the intrinsic abundances and  
indicates that the [Fe/H] ratio should not be used 
as a metallicity indicator.
An example of application of the revised dust correction procedure
in DLAs is  presented in the next section.

\begin{figure} 
\plottwo{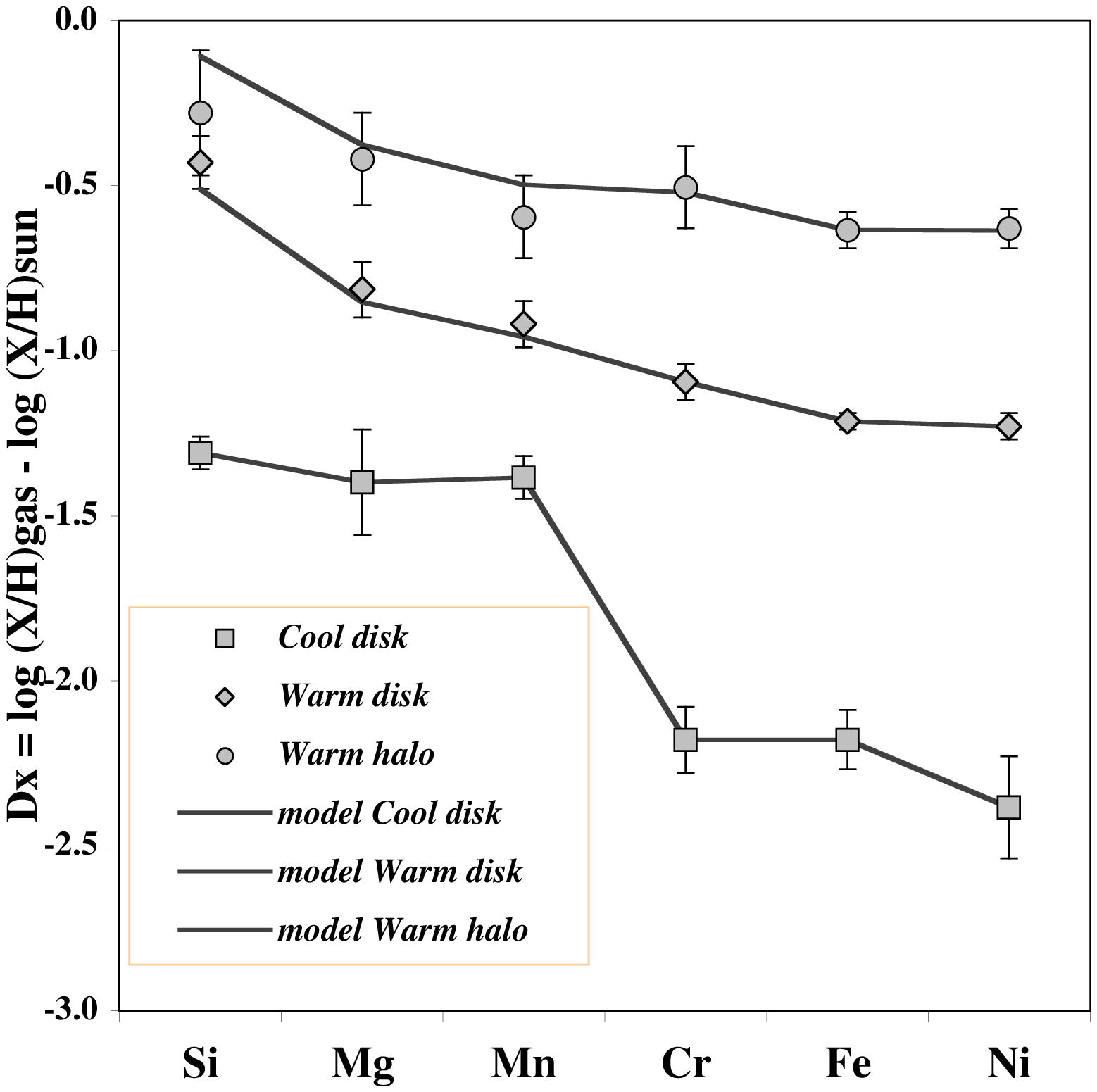}{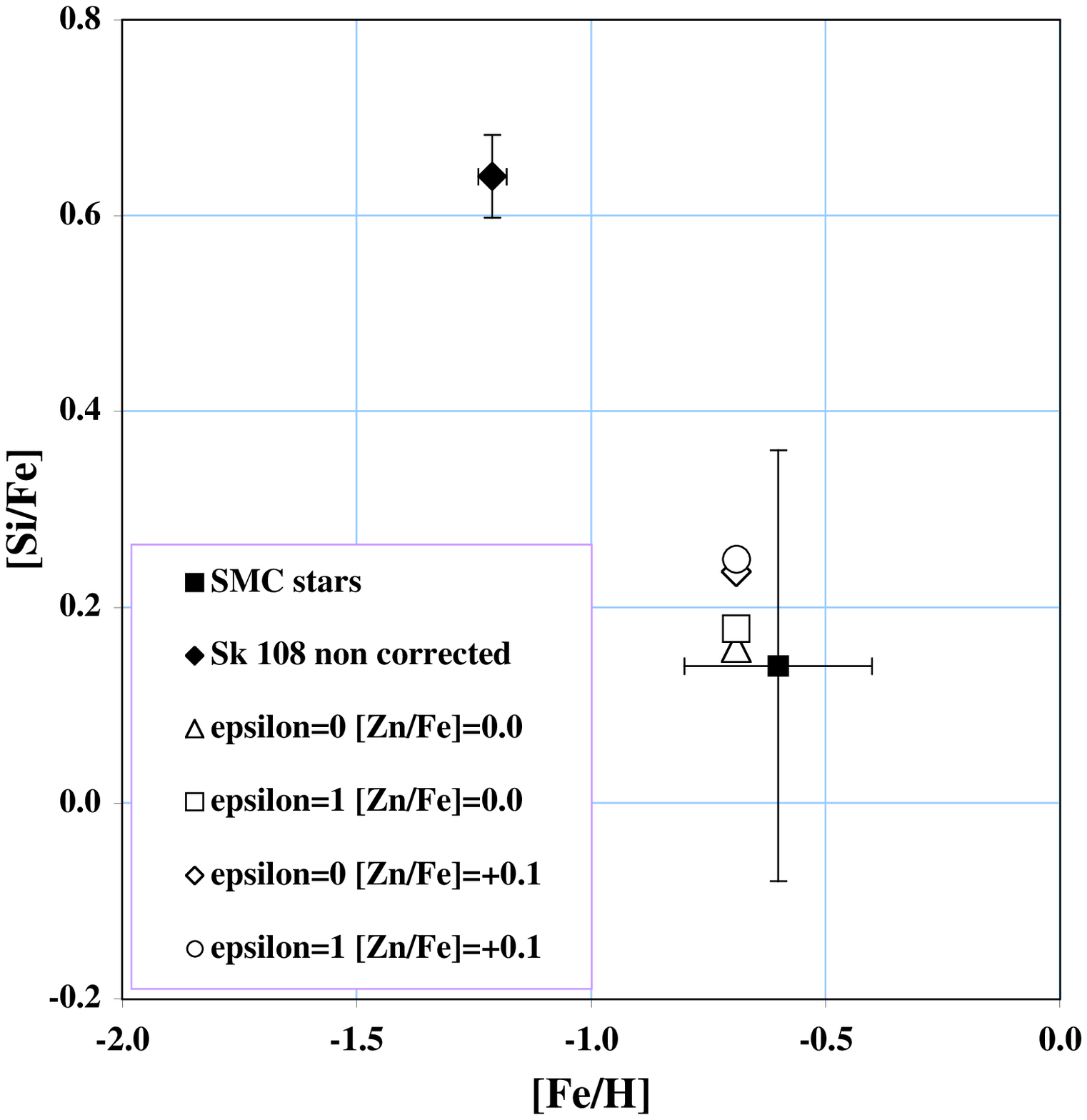}
\caption{
{\bf Left panel}: observed and predicted Milky Way depletions.
Symbols: observed depletions in the cold disk (squares), warm disk (diamonds)
and warm halo (circles)  (Savage \& Sembach 1996). 
Lines: depletion patterns modeled with Eq. (1).  
{\bf Right panel}: dust correction of the [Si/Fe] ratio
measured in the SMC interstellar gas towards Sk 108 (Welty et al. 1997).    
Filled diamond:  [Si/Fe] versus [Fe/H] not corrected for dust.
Empty symbols: dust-corrected [Si/Fe] ratios plotted versus
[Zn/H] (see legend).  Filled square: range of [Si/Fe] and [Fe/H] measurements 
typical of SMC stars (Russell \& Dopita 1992).   
  }
\end{figure}

\section{The $\alpha$/Fe ratio in DLAs}

Observations of metal-poor stars of the Milky Way indicate that
the ratio of $\alpha$-capture over iron-peak elements 
is overabundant ([$\alpha$/Fe] $\approx$ +0.5 dex at [Fe/H] $\simeq -2/-3$ dex) 
and that the overabundance declines with increasing metallicity
(Ryan, Norris \& Beers 1996 and refs. therein). 
This behaviour of the $\alpha$/Fe ratio is  interpreted in terms of the
so-called {\em time-delayed model}, based on different timescales of
production of  $\alpha$-capture over iron-peak elements  by SNe II and SNe Ia
(Matteucci 2000 and refs. therein).
The $\alpha$/Fe ratio is therefore an indicator of chemical evolution
that can also  be applied to infer  the evolutionary state of DLA galaxies. 
Here we briefly discuss the [Si/Fe] ratio,
the most commonly measured $\alpha$/Fe ratio in DLAs,
taking into account the effects of ionization and dust. 

The ionization effects act
in such a way that [Si/Fe] tends to be overestimated in DLAs and
a negative correction must be applied to recover the intrinsic
abundance (Fig. 1, right panel). The effect, however, is very small 
independent of the  ionizing continuum adopted, provided the gas sampled
along the line of sight is chemically homogeneous. Therefore
 we consider [Si/Fe]
ionization corrections to be negligible
in the rest of this discussion. 
Dust depletion yields an   enhanced [Si/Fe] ratio in the local ISM
(Savage \& Sembach 1996). 
As discussed in Section 3, dust depletion can significantly affect the
Si/Fe in a metal-poor galaxy such as the SMC (Fig. 2,right panel)
and is likely to influence the measurements in DLAs. 
The original claim  that [Si/Fe] is enhanced in DLAs 
due to nucleosynthetic processes (Lu et al. 1996)
must be therefore taken with extreme caution. 
In Fig. 3 we show an updated collection of [Si/Fe] measurements in DLAs corrected for
dust  with the procedure outlined in the previous section. 
Two conclusions can be derived from a quick look at the dust corrected
[Si/Fe] data shown in the figure: 
(1) the decline with increasing metallicity expected 
by the {\em time-delayed model} is not detected at a significant level
of statistics,
even though the dashed line resulting from the
linear regression through the data is consistent with 
the existence of such a trend; 
(2)  most of the DLAs ratios lie below the typical [Si/Fe]
values measured in metal-poor Galactic stars (continuous line). 
Both  conclusions are fairly independent of  the 
parameters adopted in the dust correction procedure (see caption to Fig. 3). 
In addition, both conclusions are supported by a study of
 the [S/Zn] ratio, a  dust-free indicator of the $\alpha$/Fe ratio
in DLAs (Centuri\'on et al. 2000).
The consistency of the results found from dust corrected [Si/Fe] ratios
and dust-free [S/Zn] ratios is an   argument in favour of
the general accuracy of the dust correction procedure. 
Independent evidence of non enhanced $\alpha$/Fe ratios
at very low metallicity is found in the dust-free
 system at $z=3.3901$ towards Q0000-26  (Molaro et al. 2000). 

\begin{figure} 
\plotone{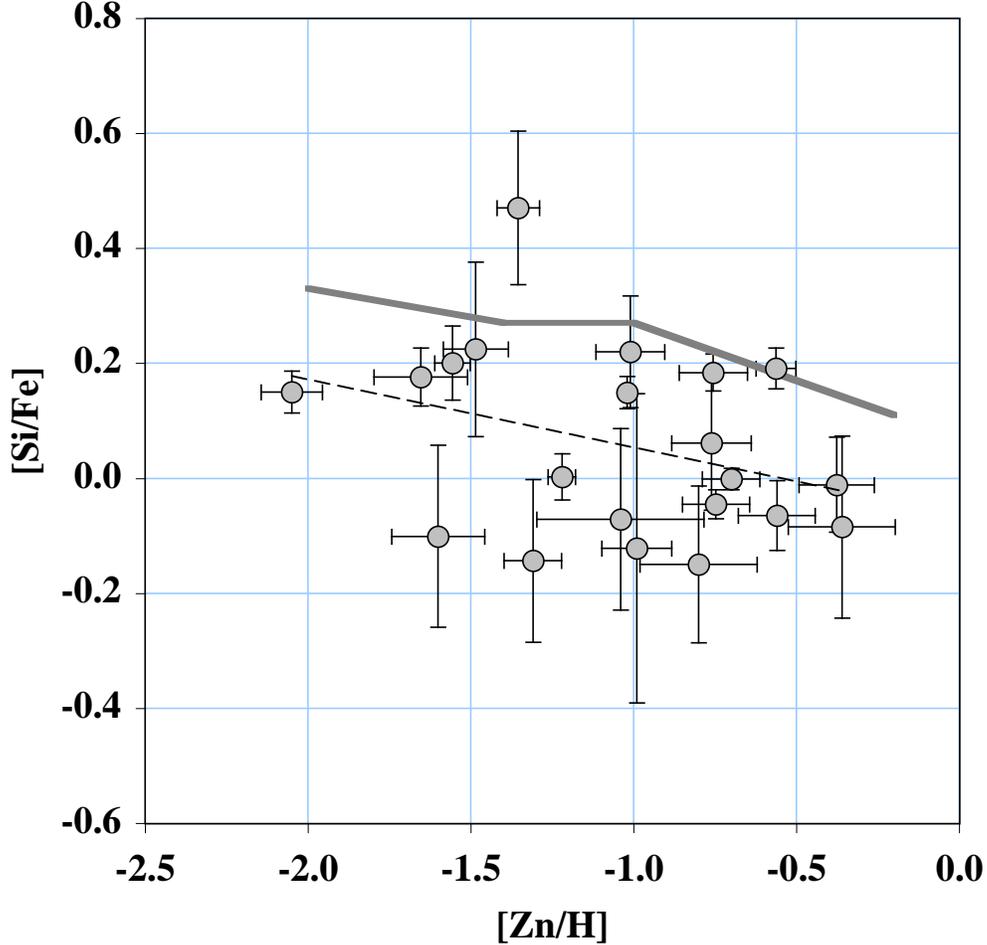} 
\caption{
Compilation of available [Si/Fe] and [Zn/H] 
 measurements in DLAs   corrected for dust depletion  
as explained in Section 3. 
Most of the original data  are taken from
Prochaska \& Wolfe (1999); the most enhanced [Si/Fe] ratio
at [Zn/H] = $-1.4$
corresponds to the $z=3.025$ absorber towards Q0347-38
(Levshakov et al. 2001); a complete list of
references will be presented in a subsequent work. 
Dotted line: linear regression through the corrected data. 
Thick curve: midmean vector of the [Si/Fe] measurements in Galactic
metal-poor stars defined by Ryan et al. (1996). 
The parameters adopted in the dust correction procedure are  
$\varepsilon_x=1$ and [Zn/Fe]$_{\rm int}$=0.0. 
By varying the input parameters  
in the ranges $0 \leq \varepsilon_{\rm x} \leq 1$ and  
$ 0.0 \leq \rm{[Zn/Fe]}_{\rm int} \leq +0.1$ the [Si/Fe] corrected
ratios  still lie below the thick curve and show a hint of
a decline with increasing [Zn/H].  
}
\end{figure}

The lack of a clear trend  of the $\alpha$/Fe ratio with increasing
metallicity is not too surprising 
considering that DLAs are probably associated to galaxies of different morphology 
and evolutionary status. We note, however, that
the modest decrease with metallicity, although not statistically significant,
is seen both in [Si/Fe] and [S/Zn]. In addition, the linear regression of the same
data versus redshift suggests a modest decrease
of both [Si/Fe] and [S/Zn] ratios with cosmic time.  
These trends with redshift, although not statistically significant, 
are also consistent with the general expectations of the {\em time delayed model}. 

The relatively low  $\alpha$/Fe values found in DLAs compared to 
those measured in Galactic stars of similar metallicity indicates 
that  DLAs have undergone a chemical evolution different from that of
the solar neighbourhood. 
The lower $\alpha$/Fe ratios at a given metallicity  
can be explained by chemical evolution
models with low or intermittent star formation rates (Matteucci 1991). From this
kind of argument an origin of DLAs in low-mass galaxies seems more
appropriate than an origin in progenitors  of galaxies as massive as the Milky Way.
This conclusion challenges the original idea that DLAs are progenitors of massive rotating
disks.
It is important to remark, however, that some DLAs do show evidence
for a  $\alpha$/Fe enhancement, the most clear example being the 
$z=3.025$ absorber towards Q0347-38 
(Levshakov et al. 2001), also shown in Fig. 3.  
 This counter-example indicates that DLAs include galaxies
with different chemical evolution and therefore that
they do not represent a homogeneous class of galaxies.

\section{Conclusions}

Abundance studies  provide unique information on the nucleosynthetic
processes and chemical evolution at work
in DLAs provided we are able to disentangle dust and ionization effects.
Ionization corrections could be large  if the bulk of the neutral gas is metal
poor while the metal lines originate in a HII region. 
This might happen in specific cases, but is probably not the rule. 
In general,  the
ionization corrections
are expected to be small and to become even  smaller
with increasing $N$(HI) for most of the elements  
commonly measured.  
On the other hand, dust depletion can significantly affect DLAs
abundance determinations. 
Exceptions are the non-refractory elements, such as N, O , S and Zn. 
While some information can be gathered from the analysis of these
elements, the study of the overall abundance pattern requires
dust corrections to be applied. 
A revised method for dust correction  has been briefly outlined in
this presentation. This method
allows the dust composition to vary 
and as a function of the dust-to-gas ratio and, for the first time,
as a function 
of the   abundance of the medium
(dust plus gas). 
This dust correction method has been successfully
tested in the interstellar gas of the SMC and applied
to correct Si/Fe measurements in DLAs. 
The dust-corrected [Si/Fe] ratios  
suggest that most DLAs have undergone a chemical history different
from that observed in metal-poor stars of the solar vicinity. 
This result challenges the concept that DLAs are progenitors
of present-day spiral galaxies. An origin in galaxies or proto-galaxies
with lower or intermittent rates of star formation is favoured
by the $\alpha$/Fe corrected data. The same data indicate
that   DLAs do not represent a homogeneous class of objects
since they include
galaxies with different types of chemical evolution.
This conclusion is consistent with the well known inhomogeneity of 
the   DLAs
galaxies identified at $z \la 1$. 
The decline of the $\alpha$/Fe ratio   expected
by {\em time-delayed} models of chemical evolution
is not detected clearly, probably as a consequence
of the inhomogeneity of the  sample. 

\acknowledgments
 I would like to thank Francesca Matteucci
and the whole  organising committee for their invitation to this
enjoyable meeting.

\end{document}